\documentclass[letterpaper,12pt]{article}
\usepackage{amsmath,amssymb}
\usepackage{graphicx}
\usepackage{courier}
\usepackage{bm}
\usepackage{pbox}
\usepackage{natbib}
\usepackage[font=small,labelfont=bf]{caption}
\usepackage[left=2.54cm,right=2.54cm,top=2.54cm,bottom=2.54cm]{geometry}
\usepackage[hyphens]{url}
\usepackage{adjustbox}
\usepackage{setspace}
\usepackage{amsthm}
\usepackage{amsmath}
\newtheorem{assum}{Assumption}

\DeclareMathOperator*{\argmin}{argmin}
\DeclareMathOperator{\sign}{sign}

\title{Wind-robust sound event detection and denoising for bioacoustics}
\author{Julius Juodakis\thanks{School of Mathematics and Statistics, Victoria University of Wellington, New Zealand. Corresponding author: Julius Juodakis, julius.juodakis@sms.vuw.ac.nz} \and Stephen Marsland\footnotemark[1]}
\date{}

\begin{document}
\bibliographystyle{rss}

\maketitle

\section*{Abstract}
\begin{enumerate}
  \item Sound recordings are used in various ecological studies, including wildlife monitoring by acoustic surveys.
Such surveys require automatic detection of target sound events in the large amount of data produced.
However, current detectors, especially those relying on band-limited energy as the main feature, are severely impacted by wind, which causes transient energy increases.
The rapid dynamics of this noise invalidate standard noise estimators, and no satisfactory method for dealing with it exists in bioacoustics, where simple training and generalization between conditions are important.

  \item We propose to estimate the transient noise level by fitting short-term spectrum models to a wavelet packet representation.
This estimator is then combined with log-spectral subtraction to stabilize the background level.
The resulting adjusted wavelet series can be analysed by standard energy detectors.
We use real data from long-term monitoring to tune this workflow, and test it on two acoustic surveys of birds.
Additionally, we show how the estimator can be incorporated in a denoising method to restore sound.

  \item The proposed noise-robust workflow greatly reduced the number of false alarms in the surveys, compared to unadjusted energy detection.
As a result, the acoustic survey efficiency (precision of the estimated call density) improved for both species.
Denoising was also more effective when using the short-term estimate, whereas standard wavelet shrinkage with a constant noise estimate struggled to remove the effects of wind.

  \item In contrast to existing methods, the proposed estimator can adjust for transient broadband noises without requiring additional hardware or extensive tuning to each species.
It improved the detection workflow based on very little training data, making it particularly attractive for detection of rare species.
\end{enumerate}

\section{Introduction}

In recent years, acoustic surveys based on long-term recordings have emerged as a powerful tool in ecology.
Such surveys can cover large scales in both time and space, making them invaluable for monitoring animal species in conservation and behaviour research (see reviews by \citet{shonfield_autonomous_2017, SugaiRev2018}).
Many further applications for such monitoring at the human-wildlife interface have been proposed, such as poaching detectors \citep{Astaras2017}, warning systems for elephant approach \citep{Zeppelzauer2014}, or farm animal welfare monitoring \citep{Mcloughlin2019}.

A key step in most of these tasks is the detection of target sounds in the recordings.
The resulting annotations can then be used in various inference models, population size estimation \citep{Dawson2009}, source localization \citep{Rhinehart2020}, or for other purposes.
Since the amounts of data recorded often total in the thousands of hours, and calls are intermittent within them, automatic detection is necessary, and choosing the right methods can have a large impact on survey efficiency \citep{acousticsurveys}. Thus, developing detectors that can be applied to natural soundscapes is an important and active area of research.

A major obstacle for current bioacoustic sound detectors is environmental noise, in particular wind \citep{NPreview}.
Wind interaction with microphones creates noise in the form of transient peaks, with higher power in lower frequencies \citep{windrevbook, NelkePhd}.
Detection in bioacoustics, at least in initial stages, typically identifies sound events as increases in energy, possibly band-filtered (e.g., \citet{Prince2019}), transformed \citep{wavelets2020} or in the spectrogram representation \citep{LasseckMC}.
Wind peaks can appear as such increases, and therefore create false positives, thus greatly reducing the detection performance.
More complex recognisers are also impaired by wind: \citet{digbyPaper13} used a decision tree based on handcrafted species-specific features that performed considerably worse in windy conditions, while \citet{Znidersic2021} observed similar issues when estimating call counts based on acoustic indices of 1-minute frames.
While the exact mechanism of this effect is not clear, rapid changes in background energy and degradation of target sound features are likely causes, and methods robust to these factors are needed to allow detection in natural conditions.

Various approaches to wind noise suppression have been developed for different tasks.
Classic denoising methods such as the Wiener or MMSE filters are not applicable to wind because of its rapid dynamics.
Neural networks have been successfully used for speech denoising, e.g., \citet{Keshavarzi2018}, and in public competitions \citep{BirdClef19}. However, their adoption in bioacoustic practice has been limited, primarily because they require large quantities of training data, which is rarely available for wildlife.
In addition, the black box nature of deep learning makes it unclear if such models would generalize to different surveys, as similar geophonic noise sources in different areas can have different noise profiles \citep{hardrain}. For example, in a recent study \citet{Vickers2021} observed that denoising by neural networks does not help subsequent call detection with unseen types of noise. Subsequent ecological inference often also makes some assumptions about the detection probability (e.g., smooth decrease with distance, \citet{Dawson2009}) that are difficult to verify with such methods, so more transparent wind-robust detectors are needed.

Some simpler methods for wind denoising have been developed in other fields, but are not applicable to bioacoustics. For example, the signal centroids method \citep{NelkeCentr} relies on the target having high dominant frequency, which is simply not true for many vocalizing species. Other methods require pitch estimation \citep{NelkePitch}, which is itself a complex task for distant and noisy sounds in natural environments.
Another distinct research area is noise mitigation by shielding, mechanical integration, or multi-microphone coherence \citep{windrevbook}.
We will not consider these options in this study, as they require physical modifications to hardware, complicate recorder deployment, and do not help analyse historical or ongoing survey data.

Therefore, we propose a new procedure for single-microphone estimation of transient broadband noise.
We use it to improve the noise-robustness of an acoustic event detection method. We will first describe the theoretical basis of this method, and then demonstrate its usage on two surveys of birds.
We also show how this estimate can be incorporated in a denoising method to restore clean sound for listening or visualization.
The proposed noise estimator is found to considerably improve the efficiency of acoustic surveys, and is easily adaptable to different species and noise profiles.

\section{Materials and methods}

\subsection{Overview of the proposed detector}

The main method proposed in this paper is a wind-robust energy detector. It detects signals in a target frequency band using these steps:
\begin{enumerate}
  \item Sound is converted to a wavelet packet tree (WPT) representation, and node(s) corresponding to the target frequencies are chosen;
  \item The noise level in the chosen node is interpolated using a log-log line, fitted to the energies in non-target nodes by least-squares or quantile regression;
  \item The estimated noise level is used in log-scale spectral subtraction to adjust the target band energy;
  \item Adjusted energies are analysed by a changepoint detection algorithm, presented previously \citep{soundchps21}, to detect increases, which are assumed to be calls.
\end{enumerate}

We will now present each of these components in more detail, starting with the final detection stage which is used to guide the other parts of the method.

\subsection{Energy-based signal detection}

In the energy detection framework, the sequence of observations $Y_t$ is modelled as the sum of a stationary noise process $N_t$ and signal $S_t$:
\begin{equation}
  Y_t = N_t + S_t.
  \label{eq:mained}
\end{equation}
The signal is transient, and its presence is detected based on the observed energy $y^2_t$.
This is motivated by assuming that both $N,S$ are independent white Gaussian processes:
\begin{equation}
  N_t \sim \mathcal{N}(0, \sigma^2_N), ~ S_t \sim \mathcal{N}(0, \sigma^2_S(t)).
  \label{eq:bsmodel}
\end{equation}
Then, testing $y^2_t$ against a fixed threshold is a generalized likelihood ratio test for the hypotheses:
\begin{align*}
  H_0: \sigma^2_S(t) = 0  \\
  H_1: \sigma^2_S(t) > 0
\end{align*}
and its properties, such as error rates, can be determined theoretically \citep{Chen2010}. For example, false alarm rates can be controlled at rate $\alpha$ by setting the threshold to $\lambda = \sigma^2_N F^{-1}(1-\alpha)$, where $F$ is the CDF of $\chi^2_1$ distribution, and $\sigma^2_N$ is estimated utilising the stationarity assumption, e.g., from quiet frames.
Larger intervals $[s,e]$ can be tested using windowed statistics such as $\sum_{i=s}^e y^2_i$ or $\max_{i \in [s,e]} y^2_i$.
The energy can also be compared at many positions, to locate the start and end of signal activity (changepoint detection; \citet{Page1954}). We will use a variation of this changepoint procedure, presented in \citet{soundchps21}, as the main detector in this study, but our results apply to any method that uses the energy $Y^2_t$ for detection.

\begin{figure}[h]
  \centering
  \includegraphics[width=0.9\textwidth]{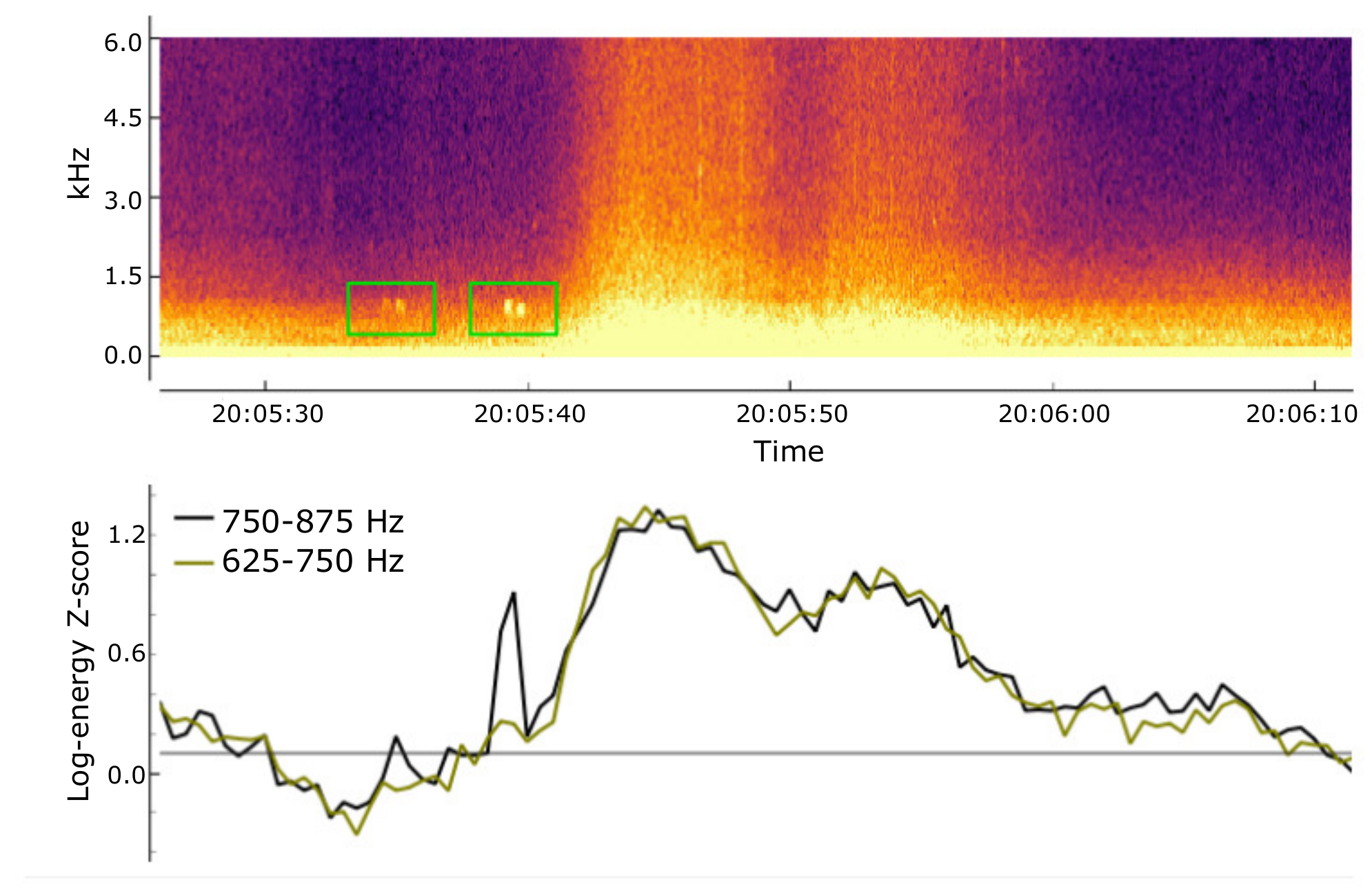}
  \caption{Top: a spectrogram showing two calls of a morepork (New Zealand owl, \emph{Ninox novaeseelandiae}) surrounded by wind noise. Bottom: wavelet coefficient log-energy for two frequency bands, extracted from the same clip in 0.5 s windows and standardized. The target band (black) shows two energy peaks corresponding to the calls, but a threshold that would capture these peaks (marked in grey) is also exceeded by the wind gust seen here. However, the wind energies in this band and a non-target band (green) appear correlated, suggesting that they could be estimated by an appropriate interpolation across frequencies.}
  \label{fig:callsinwind}
\end{figure}

In all the above methods, the stationary noise model \eqref{eq:bsmodel} is key to detection.
Wind, and transient broadband noises in general, violate this assumption and harm the performance.
For example, if the noise is Gaussian with transient increases in power:
\begin{equation}
  Y_{t} = N_{t} + S_{t}, ~ N_t \sim \mathcal{N}(0, \sigma^2_N + \sigma^2(t)),
  \label{eq:winded}
\end{equation}
then the test $y^2_t \lessgtr \sigma^2_N F^{-1}(1-\alpha)$, established on quiet periods as before, will have a false alarm rate greater than $\alpha$.
In fact, without assuming any further features distinguishing the wind and signal processes, the situations $Y_{t} = S_{t}$ and $Y_{t} = N_{t}$ are not identifiable: this can be seen in an example of a (band-limited) energy series where both a bird call and a wind gust correspond to a transient increase (Figure \ref{fig:callsinwind}). Therefore, no conclusions about the performance of energy detection under these conditions can be made, in contrast to the stationary background model \eqref{eq:mained}--\eqref{eq:bsmodel}.

\subsection{Wavelet packet representation}
Energy detection is typically applied to a band-limited representation of the recorded sound, in order to reduce the impact of non-target signals.
In this paper, we will use sound frequency subbands obtained from a wavelet packet tree (WPT).
Bioacoustic detectors using WPT have been used previously \citep{Zhang2015, wavelets2020, soundchps21}.
It is a multiscale decomposition, defined at each scale $j=1,\dots,J$ by a set of orthogonal filters $\{ \psi_{j,k} \}_{k \in 1,\dots,2^j }$, which approximately bandpass the signal to the frequency range $[ F_N(k-1)/2^j, F_Nk/2^j]$. Here $F_N$ is the Nyquist frequency, equal to half of the sampling rate.
Denoting by $X_t$ the original sound waveform, each node $(j,k)$ of the WPT thus contains the series of coefficients:
\begin{equation*}
  Y_{j,k,t} = \sum_{i=1}^n \psi_{j,k}(t - i) X_{i}
\end{equation*}
obtained by applying the filter $\psi_{j,k}$.

Before application, the user selects a node $(j_0, k_0)$ likely to contain the target signal, i.e., the signal is detected in $Y_t := Y_{j_0, k_0, t}$.
The choice can be made a priori based on the expected frequency range of the target, or by a training process, for example as described in \citet{wavelets2020}.
(We will assume that a single target node is chosen for simplicity of notation, but the method proposed here directly applies with several nodes as well.)

In addition to removing out-of-band interference, the wavelet transform decorrelates various types of noise \citep{Wornell1993}, thus allowing methods derived under white Gaussian noise assumptions to be applied to more diverse problems. Increases in energy, caused by either in-band signal or transient broadband noises, are preserved by this transform (Figure~\ref{fig:callsinwind}).

\subsection{Interpolating noise level}
To allow detection in windy conditions, we propose a new method for estimating the level of transient broadband noise.
The main idea is, at each time point, to fit a regression line to subband energies, and use that to interpolate the energy in the target band.

We will assume for now that the following model consisting of only wind and target signal, and later discuss relaxing it to allow other signals and noises.

\begin{assum}
  \label{eq:as1}
For any WPT node $j,k$:
\begin{equation*}
  Y_{j,k,t} = \begin{cases} N_{j,k,t} + S_{j,k,t} & \text{if } k=k_0 \text{ (target node) } \\
    N_{j,k,t} & \text{otherwise},
  \end{cases}
\end{equation*}
where $N_{j,k,t}$ and $S_{j,k,t}$ are the wavelet-transformed wind noise and signal components, respectively.
\end{assum}

We wish to estimate $N_{j,k_0,t}$ (or the noise power $|N_{j,k_0,t}|^2$), while for all other bands the $N_{j,k,t} = Y_{j,k,t}$ are directly observed.

We start with the following formula of short-term wind power spectral density, which was derived from fluid dynamics and has since been verified empirically \citep{windrevbook, NelkePhd}:
\begin{equation}
  PSD(f, t) = c(t) / f^{\alpha},
  \label{eq:windmodel}
\end{equation}
where $f$ is the frequency, $\alpha$ a constant, and $c(t)$ a factor that depends on the wind strength at time $t$ and microphone properties.
This is an example of the $1/f$ class of processes, for which decorrelation by wavelet transform has been extensively studied:
the resulting wavelet coefficients are almost uncorrelated within and between nodes \citep{Wornell1993}.
In addition, their distribution is Gaussian if the original $1/f$ process is Gaussian, also known as fractional Brownian noise (proven trivially, by linearity of the transform), or for other processes converges to Gaussian with sufficiently large $j$ \citep{Atto2010}, or by averaging multiple coefficients \citep{Serroukh2000}.

Thus, we will continue assuming that at each time point, the wind WPT coefficients vector $\mathbf{N}_{j,t} = [N_{j,1,t}, \dots, N_{j,2^j,t} ]$ is multivariate Gaussian:
\begin{equation*}
  \mathbf{N}_{j,t} \sim \mathcal{N}(0, \mathrm{diag}(\sigma^2_{j,1}(t), \dots, \sigma^2_{j,2^j}(t))).
\end{equation*}
The variance of the coefficients is close to the PSD at each node's centre frequency $f_{j,k} = F_N (k-1/2)/2^j$ \citep{AttoWPfBm, Moulines2007}:
\begin{equation*}
  \sigma^2_{j,k}(t) \approx PSD(f_{j,k}, t) = c(t) / f_{j,k}^{\alpha}.
\end{equation*}

Their energy is the square of a Gaussian and thus $\chi^2$-distributed:
\begin{equation*}
  |N_{j,k,t}|^2  \sim c(t) f_{j,k}^{-\alpha} \chi^2_1  \Rightarrow \log |N_{j,k,t}|^2  \sim -\alpha \log f_{j,k} + \log c(t) + \log \chi^2_1.
\end{equation*}

In other words, there is a linear relationship between $\log |N_{j,k,t}|^2$ and $\log f_{j,k}$.
This suggests that the target band noise level $\log |N_{j,k_0,t}|$ could be interpolated using an OLS regression of $\mathbf{x}_k = [1,  \log f_{j,k} ]$ and $\mathbf{y}_k = \log |Y_{j,k,t}|^2$:
\begin{equation}
  \label{eq:olsinterp}
\hat{N}^{OLS} := \mathbf{x}_{k_0} \hat{\beta}, \:
  \hat{\beta} = \argmin_\beta \sum_{k \neq k_0} (\mathbf{y}_k - \mathbf{x}_k \beta)^2. 
\end{equation}

Furthermore, we can use the properties $\mathbb{E} \log \chi^2_n = \psi(n/2) + \log 2$, where $\psi$ is the digamma function, and $Var(\log \chi^2_n) = \zeta(2, n/2)$ with $\zeta$ being the generalized Riemann (Hurwitz) zeta function \citep{AbryVeitch}, to obtain:
\begin{align}
  \mathbb{E} \log|N_{j,k,t}|^2 &= -\alpha \log f_{j,k} + \log c(t) + \psi(1/2) + \log 2  \nonumber \\
  Var (\log|N_{j,k,t}|^2) &= \zeta(2, 1/2)
  \label{eq:residvar}
\end{align}
In other words, the error term is homoscedastic, so standard OLS results can be applied to show the estimation properties. In particular, we have that, under Assumption \ref{eq:as1} and wind model \eqref{eq:windmodel}, it is consistent: $\hat{N}^{OLS} \xrightarrow{p} \mathbb{E} \log |N_{j,k_0,t}|^2 $.

Note also that instead of individual coefficients, short-term sums $\log \sum_{i=s}^{e} |Y_{j,k,i}|^2 $ could be used as the $\mathbf{y}_k$ in the regression. If the wind strength factor $c(t)$ remains the same over this window, repeating the analysis above shows that the only change is smaller variance of the error term, as $\zeta(2,(e-s)/2)$ decreases for longer windows.

\textbf{Relaxing Assumption 1.}
While the $1/f$ model covers a variety of noise processes, noise spectra obtained in field conditions may deviate from this model, due to frequency response of the microphone, shielding, and the recording device.
Other noises may also be present. For example, white noise can be captured by the model as the special case with $\alpha=0$, but if both white and wind noise are present at comparable power, the resulting spectrum will no longer be $1/f$.
To allow adaptation to these issues, we propose including higher polynomial degrees of $\log f_{j,k}$ in the regression \eqref{eq:olsinterp}.
We investigate possible choices of the polynomial degree on a pilot dataset in a later section.

Alternatively, \citet{Achard2010} proposed some estimators designed to specifically reduce bias for estimation of contaminated $1/f$ processes.
However, these estimators only outperform the basic $\hat{N}^{OLS}$ if the contamination model is specified correctly, and even then require sufficient sample size, so we do not explore them further here.

Further, other narrow-band signals may be present, causing local deviations from the $N_{j,k,t}$ model. We then model the noise power as a contaminated distribution
\begin{equation}
   |N_{j,k,t}|^2 \sim \begin{cases}
    c(t)f^{-\alpha}_{j,k} \chi^2_1  \text{ with probability } 1-\epsilon \\
    Z_{j,k,t} \text{ with probability } \epsilon
\end{cases}
  \label{eq:contamdistr}
\end{equation}
where  $Z_{j,k,t}$ is a random variable representing the power of contaminating signal and noise, and $\epsilon$ is the rate of contamination.
If the contamination is sufficiently loud, so most of the mass of $Z_{j,k,}$ is above $c(t)f^{-\alpha}_{j,k}$, the distribution of $ |N_{j,k,t}|^2$ can be highly asymmetric, and thus estimation by OLS is significantly biased, especially if higher order terms are included.

An intuitive solution is to replace the square loss \eqref{eq:olsinterp} used in regression with an asymmetric loss function, such as:
\begin{equation}
  L_\tau(\beta) = \sum_{k: k \neq k_0, \mathbf{y}_k \ge \mathbf{x}_k \beta} \tau |\mathbf{y}_k - \mathbf{x}_k \beta| +
  \sum_{k: k \neq k_0, \mathbf{y}_k < \mathbf{x}_k \beta} (1-\tau) |\mathbf{y}_k - \mathbf{x}_k \beta|
  \label{eq:quantregloss}
\end{equation}
and then estimate the noise level by:
\[ \hat{N}^{Q,\tau} := \mathbf{x}_{k_0} \hat{\beta} , ~ \hat{\beta} = \argmin_\beta L_\tau(\beta). \]
This procedure is in fact identical to quantile regression (QR), so $\hat{N}^{Q,\tau}$ estimates the $\tau$-quantile of the (contaminated) noise level given $\mathbf{x}_{k_0}$ \citep{Koenker1978}.
This allows us to use the known properties of quantile regression to analyse this estimator.

It can be shown that if sufficiently low quantile $\tau$ is chosen, then $\hat{N}^{Q,\tau}$ estimates the $\tau/(1-\epsilon)$-quantile of the uncontaminated noise distribution (Supplementary Material \ref{sec:contquantilesect}). Thus the interpolation of the noise level is biased, but a bias adjustment factor can be obtained if $\hat{\epsilon}$ is available.
Alternatively, note that by averaging the energy statistic over many windows, arbitrary variance reduction of the noise can be achieved, with all quantiles approaching $\mathbb{E}(\log|N_{j,k,t}|^2)$, and thus $\hat{N}^{Q,\tau}$ can be made consistent.

\subsection{Proposed adjustment method: log spectral subtraction}


Once an estimate of the noise level is obtained, it can be used to estimate the clean signal $\tilde{Y}_t$ from $Y_t$.
A common method for this is spectral subtraction \citep{VaseghiBook}, stated in its power form as:
\begin{equation}
  |\tilde{Y}^2_t| = \max(0, |Y^2_t| - |\hat{N}^2_t|),
  \label{eq:spsub}
\end{equation}
where $\hat{N}_t$ is an estimate of $N_t$. This results in strong suppression in low SNR conditions, which is desirable for many applications.

\begin{figure}[h]
  \centering
  \includegraphics[width=0.8\textwidth]{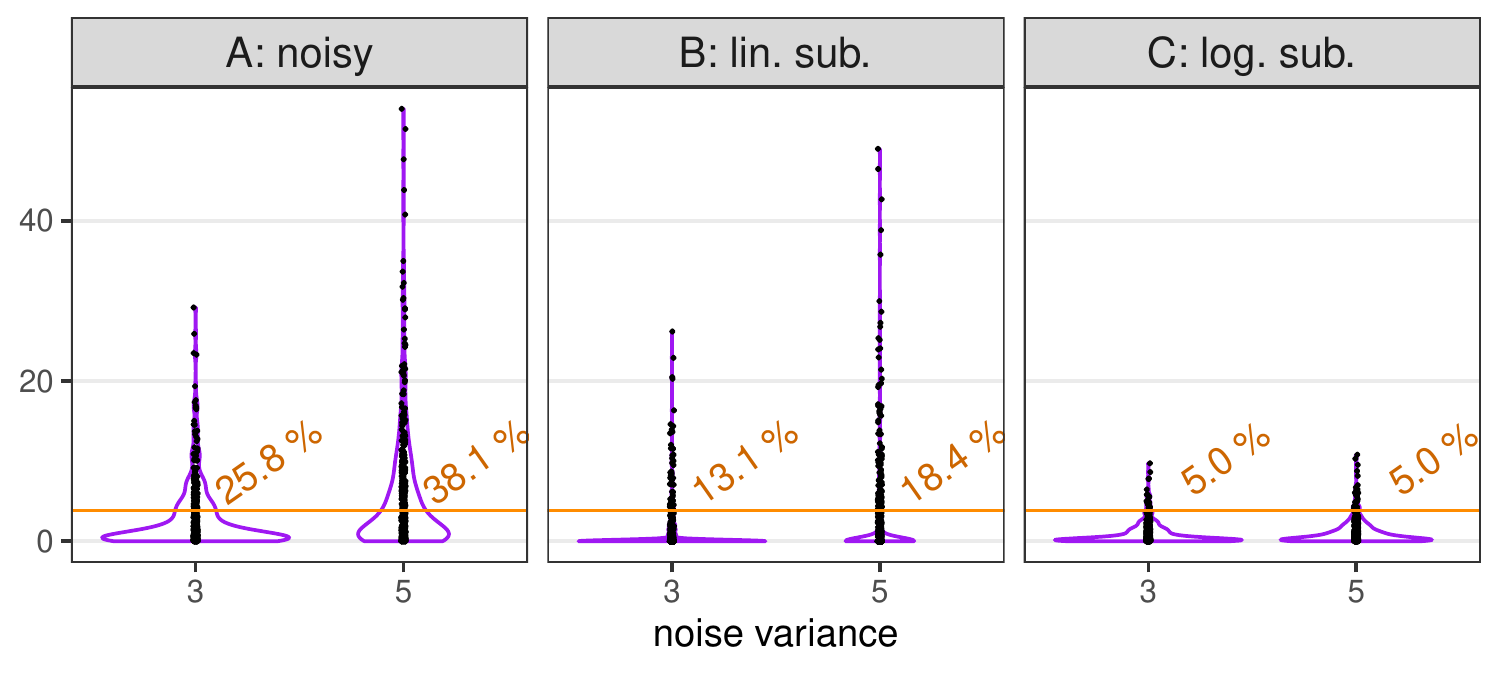}
  \caption{Violin plots showing the distribution of energy of: (A) Gaussian noise with $\sigma^2=$ 3 or 5, (B) same noise after applying spectral subtraction with the true $\sigma^2$, (C) same noise after log spectral subtraction with the true $\sigma^2$. Dots are 500 simulated points and purple lines are kernel density estimates. The numbers indicate the probability of exceeding the threshold $\lambda=3.84$ (5th percentile of $\chi^2_1$, orange line) in each case. See Supplementary Material for calculation details.}
  \label{fig:chisq}
\end{figure}

However, this adjustment does not work well for signal detection.
So far, we have assumed that the noise power $N^2_t$ is a $\chi^2$-distributed random variable, and interpolation can at best provide some estimate of its expectation $|\hat{N}_t^2| = \sigma^2_{j,k}(t)$. 
The distribution of $\tilde{Y}^2_t$, produced by spectral subtraction, will then be a left-censored and shifted $\chi^2$ distribution (Figure \ref{fig:chisq}A-B).
Furthermore, we show that this adjusted distribution will still depend on $\sigma^2_{j,k}(t)$ (Supplementary Material \ref{sec:spsubtailsect}).
Both of these issues violate the stationary Gaussian model \eqref{eq:bsmodel}, and so energy-based detectors applied after spectral subtraction even with perfect estimates will still not have the expected performance.

We propose that spectral subtraction for detection should be carried out on log scale, or equivalently:
\[ |\tilde{Y}^2_t| = \max(1, |Y^2_t| / |\hat{N}^2_t| ). \]
In contrast to \eqref{eq:spsub}, this log spectral subtraction will produce $\tilde{Y}^2_t$ distributed as $\chi^2_1$ (Figure \ref{fig:chisq}C).
Given an accurate estimate $|\hat{N}_t^2|=\sigma^2_{j,k}(t)$, the adjusted estimate will match model \eqref{eq:bsmodel} with $\sigma^2_N=1$, and optimal performance of the detection methods derived under this model can be expected.

\subsection{Validating the noise spectrum fit on pilot data}
To investigate whether field recording data matches the $1/f$ spectrum profile, we conducted a pilot experiment on a set of short clips randomly selected from a larger monitoring project.

Over 2018--2019, nightly acoustic monitoring was conducted with passive recorders in Zealandia sanctuary, Wellington, New Zealand. We selected five nights of recordings from this data, obtained over various months using two devices (SM2, Wildlife Acoustics). Recorders were attached to trees at about 1--1.5 m above ground, with one located in a relatively exposed position on a hilltop, and the other in a sheltered valley. We extracted 3 audio clips of 0.1 seconds from each night and device, starting within one minute of 23:00, and manually verified that no distinct animal calls are heard in these clips.
For comparison with a different hardware, we extracted 3 similar clips from one night from a different monitoring project, conducted in 2021 in Ponui island, New Zealand, using an AR4 recorder (Department of Conservation). All clips were resampled to 16000 Hz.

The clips were subject to WPT using two different wavelets: discrete Meyer, which approximates ideal bandpassing, and order 8 Symlet, based on its estimation performance in \citet{AttoWPfBm}.
Energy within each node ($|Y_{j,k,t}|^2$ using the notation above) was averaged over the 0.1 s clip and shown as the spectrum estimate at node centre frequency $f_{j,k}$.
For comparison, we also plotted the spectra obtained from the clips by periodogram, Daniell-smoothed with a 7-bin kernel and downsampled.

OLS regression models were then fitted to the log frequencies vs. log energies, as in \eqref{eq:olsinterp}. We used either all nodes between 150--7500 Hz (``full spectrum'', excludes only edge bands that have filtering effects) or nodes between 150--6000 Hz, to focus on the more likely wind range.
A series of models were fitted, from linear to 6th degree polynomial, and evaluated by the small-sample corrected Akaike criterion: $AICc = -2 \log L + 2kn/(n-k-1)$.

\subsection{Case study: applying the proposed noise-robust detection}
We demonstrate the proposed wind-robust detection method on two bird surveys.
The first is a survey of Australasian bittern (\textit{Botaurus poiciloptilus}), conducted near Lake Ellesmere, New Zealand. The male bitterns emit `boom' calls at low frequency (around 150 Hz), meaning that their detection is particularly affected by wind. The survey was conducted for 2 hours using 7 recorders at 8 kHz sampling rate. Playback was used to solicit calls, and for the purpose of method evaluation we count both playback and responses as true calls.
The second survey is of little spotted kiwi (\textit{Apteryx owenii}) in Zealandia wildlife sanctuary, New Zealand. Male kiwi calls are a sequence of around 20 repeated syllables in the 2-3 kHz band.
Eight recorders with 6 hours of sound from each were used.
These surveys were previously used to evaluate sound detection methods in \citet{soundchps21}, and further details about this data are provided therein.

The recordings were analysed using the changepoint detector from \citet{soundchps21}. Briefly, a training process uses a small number of annotated files to characterize the wavelet nodes and duration of each species calls.
The survey files are then analysed to detect periods of increased energy in these nodes. The detector can adapt to long-term changes in background level, but transient events such as wind are not removed and cause false positives \citep{soundchps21}.
The wind-adjusted analyses use the same detectors with the same parameter settings, but reduce noise level by log-spectral subtraction as described here. The analysis was repeated using either the OLS or QR noise estimate. The same window length is used for both detection and fitting of the noise spectrum models.
For the quantile estimate $\hat{N}^{Q,\tau}$, we set $\tau=0.2$, and in the case of bittern the estimate was adjusted upwards by 0.4 (this factor, based on \eqref{eq:residvar}, is typically negligible and only used here because of the low sampling rate of the bittern recordings).

Evaluation is based on the precision of a spatial capture-recapture model (SCR), as proposed previously in \citet{acousticsurveys}.
SCR is a general framework for inferring population density from imperfectly detected cues \citep{Dawson2009}.
Its key component is a detection function $p(d)$, modelling the probability of detecting calls emitted at distance $d$ from the recorder.
In the grid-based SCR, as used here, this probability is estimated from calls simultaneously detected by more than one recorder.
Another option is to calibrate $p(d)$ from external data, in which case the SCR reduces to the distance sampling model \citep{borchersunifying2015}.
The density of animal calls, assumed proportional to the density of animals, is estimated using this $p(d)$.
As this density is the main target of ecological interest, we use its standard error (SE) to evaluate the detection methods.

After applying the detection algorithms, equal number of reported segments from each method were reviewed manually. The verified detections were used to fit an SCR model, and the density SE estimated by bootstrap \citep{stevenson_general_2015} is reported, as well as the coefficient of variation $SE(\hat{D})/\hat{D}$ to allow differences in the density estimate $\hat{D}$.
We refer the reader to \citet{borchersunifying2015, stevenson_general_2015} for a full introduction to acoustic SCR, and to \citet{acousticsurveys} or \citet{soundchps21} for details on formatting data for this type of model.

\subsection{Using the noise estimate to restore clean sound}

The proposed estimator of broadband noise level ($\hat{N}^{OLS}$ or $\hat{N}^{Q,\tau}$) can also be combined with other sound analysis methods, not only detectors. We demonstrate how it can be used for restoring clean sound by wavelet shrinkage.

Wavelet shrinkage by soft-thresholding is a popular denoising method \citep{Donoho1995}.
The soft-thresholding modifies the WPT coefficients $Y_{j,k,t}$ by translating them towards 0:
\begin{equation}
  \tilde{Y}_{j,k,t} = \sign(Y_{j,k,t}) \max(0, |Y_{j,k,t}| - \lambda \sigma_{j,k,t})
  \label{eq:wvshrink}
\end{equation}
where $\sigma_{j,k,t}$ is some estimate of the noise SD in the node $j,k$, and $\lambda$ tunes the strength of the thresholding.
This is based on the assumption that target signal energy will be concentrated in only a few coefficients after the wavelet transform, and so shrinking all coefficients will mostly reduce noise.
The adjusted WPT is then inverted to reconstruct a denoised sound waveform (see e.g., \citet{Wornell1993}), which is simpler compared to inverting a spectrogram.

As a test, we create noisy files by mixing 2 min clips of windy background with bird sound examples.
Background clips (5 files) were selected from Zealandia monitoring data. Bird sounds were 6 clips from the xeno-canto database and 6 clips of rich soundscapes from Zealandia monitoring. The xeno-canto examples were chosen to have a clear foreground and low background noise, because evaluating the denoising requires clean reference sounds. The Zealandia examples were taken from dawn or dusk choruses, to capture rich soundscapes that are difficult to denoise, although they have non-negligible background noise, which may impact the subsequent denoising metrics.
The clips were mixed at +12 dB, 0 dB, or -12 SNR (the latter was only used with xeno-canto examples, as the soundscapes are too quiet to produce audible residual signal then), producing 300 min of noisy sound in total.

Each file was then analysed by constructing the WPT, and for each time window noise level estimates $\hat{N}^{OLS}, \hat{N}^{Q,\tau}$ were obtained by fitting a cubic polynomial to the WPT as described above.
Since these values estimate the log-energies of noise, we can obtain adaptive estimates of $\sigma_{j,k,t}$ as $\sqrt{\exp(\hat{N}^{OLS})}$ or $\sqrt{\exp(\hat{N}^{Q,\tau})}$. We use these and $\lambda=1$ in \eqref{eq:wvshrink} to obtain OLS-denoised or QR-denoised coefficients.
For comparison, we test a constant threshold with $\lambda=3$ and $\sigma_{j,k,t} = \sum_{i=1}^{n}|Y_{j,k,i}-\mathrm{median}(Y_{j,k,i})| / (0.6745n)$; this is a robust estimate of the noise SD, commonly recommended for wavelet shrinkage, and leading to various optimal theoretical properties \citep{Donoho1995}.
The resulting adjusted WPT was inverted to reconstruct the denoised sound file following standard wavelet methods \citep{Donoho1995}.

The success of denoising was evaluated by estimating the SNR improvement in dB:
\[ SNR_{impr.} = SNR_{denoised} - SNR_{noisy} =10 \log_{10} \sum_t X_t^2/(X_t-\hat{X}_t)^2 - 10 \log_{10} \sum_t X_t^2/(X_t - X^N_t)^2, \]
where $X_t, X^N_t$ and $\hat{X}_t$ are the clean, noisy and denoised waveforms of a file.
We also calculate the SI-SDR, which is a robust modification of SNR that is invariant to scale changes introduced during denoising \citep{Roux2019}.

\section{Results}
\subsection{Field recordings indicate non-linear background spectra}

\begin{figure}[h]
  \centering
  \includegraphics[width=\textwidth]{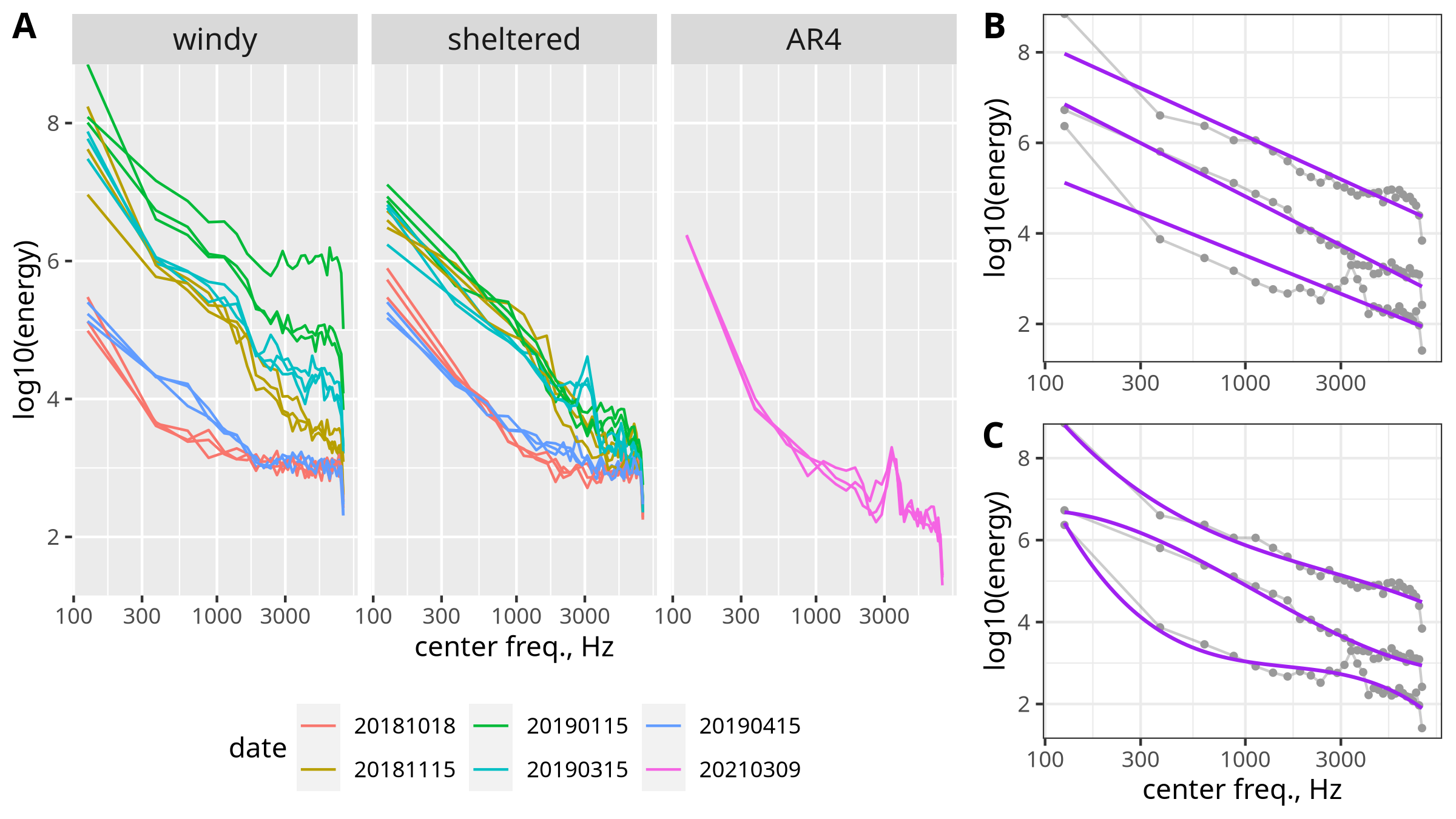}
  \caption{(A) Short-term spectra (energies in discrete Meyer wavelet packet nodes) of 0.1 s clips from passive acoustic recordings. The log energy of each node is plotted against its centre frequency. Panels show clips from a recorder in a windy location, a sheltered location, and a different model recorder (AR4). (B) A subset of three example spectra (grey connected points) overlaid with linear fits. (C) Same example clips with cubic fits.}
  \label{fig:pilotres}
\end{figure}

The pilot dataset revealed the presence of a variety of background noise spectra (Figure \ref{fig:pilotres}A).
Overall, the noise power was higher in lower frequencies, higher in the January and March nights when more wind gusts were audible, and more variable for the recorder in an exposed location, which is in line with the $1/f$ model (note that some files have an additional peak corresponding to strong cicada noise, at around 3000 Hz).
Spectra taken close in time to each other show little variation, suggesting that the estimation is precise in stable conditions.

However, over longer periods, spectral shapes varied considerably, deviating from the predicted log-log line.  Even within the same device and same minute, wind gusts caused some considerable changes in spectrum shape (see top lines in Figure \ref{fig:pilotres}A, ``windy'' recorder).
Similar spectra were obtained with a different wavelet, or by a smoothed periodogram (Supplementary Material, Figure \ref{fig:pilotressupp}), indicating that the shape variation is not caused by the chosen estimation method.

Linear models were also not supported by the fit statistics: when fitting the full-spectrum, average AICc for the linear model was 35.0, while the higher order polynomials had AICc between 17.0--23.6. Even if the frequency range is limited to \textless 6000 Hz, the linear model is still insufficient (AICc 28.5, but 16.5--21.3 for higher order models). The optimal model degree by this criterion was 5 (for full spectrum) or 3 (for \textless6000 Hz).
Some examples of 3rd degree and linear spectrum fits are shown in Figure \ref{fig:pilotres}B-C. Based on these results, we chose to use a 3rd degree model fitted to \textless 6000 Hz spectrum in the detector, as it seems to provide sufficient flexibility in the range where wind noise is the most prominent, without great sensitivity to interferences or large computational cost.

\subsection{Evaluating robust detection on surveys}

In all tested settings, we observed that wind adjustment greatly reduced the number of false positives.
In the bittern survey, 859 detections were obtained using the OLS-adjusted method. Without the adjustment, 1505 segments were reported, of which 57\% were reviewed to equalize the effort across both methods.
Most of the additional false positives in this set were indeed wind or other broadband noises such as plane overflights, which were also removed by the proposed adjustment.
Fitting SCR models to the two sets of detections confirms that the adjustment greatly improves survey efficiency, with about two-fold lower coefficient of variation for the estimated density (Table \ref{tab:scrres}).

An even greater contrast is seen in the kiwi survey. With the same thresholds, the adjusted detection resulted in so few false positives that it required extreme downsampling of the unadjusted data (SCR models could not be reliably fitted).
Therefore the results shown here use a two times smaller threshold for the adjusted detector. This produced 323 detections, mostly true positives, with the rest caused by sounds of other species in the target bands, such as the kaka parrot (\textit{Nestor meridionalis}).
In comparison, the unadjusted detector produced 1315 detections (25\% reviewed), and 4 times less precise density estimates (CoV 52.5\% vs 12.2\%, Table \ref{tab:scrres}).

\begin{table}[h]
  \caption{\label{tab:scrres}Detection results from two bird surveys, obtained using a wavelet changepoint detector with or without a wind noise adjustment. The adjustment uses the OLS spectrum fit presented in this paper. The main evaluation metrics are highlighted: standard error (SE) or the coefficient of variation (CoV) of the survey density estimate. Also shown are the estimates of the density itself and of the detection radius parameter $\sigma$.}
  \centering
  \begin{tabular}{l|rr|rr}
    \hline \hline
    & \multicolumn{2}{c|}{Bittern} & \multicolumn{2}{c}{Kiwi} \\
    & no adj. & OLS adj. & no adj. & OLS adj. \\ \hline
    Total detections & 1505 & 859 &   1315 & 323 \\
    $\sigma$ (m)     &  329 &  263      &  290 &  232 \\
    Density (calls/ha) &  0.36 &  0.52    & 1.08 &  1.03 \\
    \textbf{Density SE} &  0.35 &  0.27    & 0.55 &  0.13 \\
    \textbf{Density CoV (\%)} &  99.5 &  \textbf{52.5}   & 51.1 & \textbf{12.2} \\
    \hline \hline
  \end{tabular}
\end{table}

When estimating wind noise by quantile regression instead of OLS, slightly more false positives were produced, with 1025 total detections for bittern and 360 for kiwi.
The detected segments mostly matched those reported by OLS, so we do not analyse these further here.

\subsection{Incorporating the wind estimator into denoising}

\begin{figure}[h]
  \centering
  \includegraphics[width=0.8\textwidth]{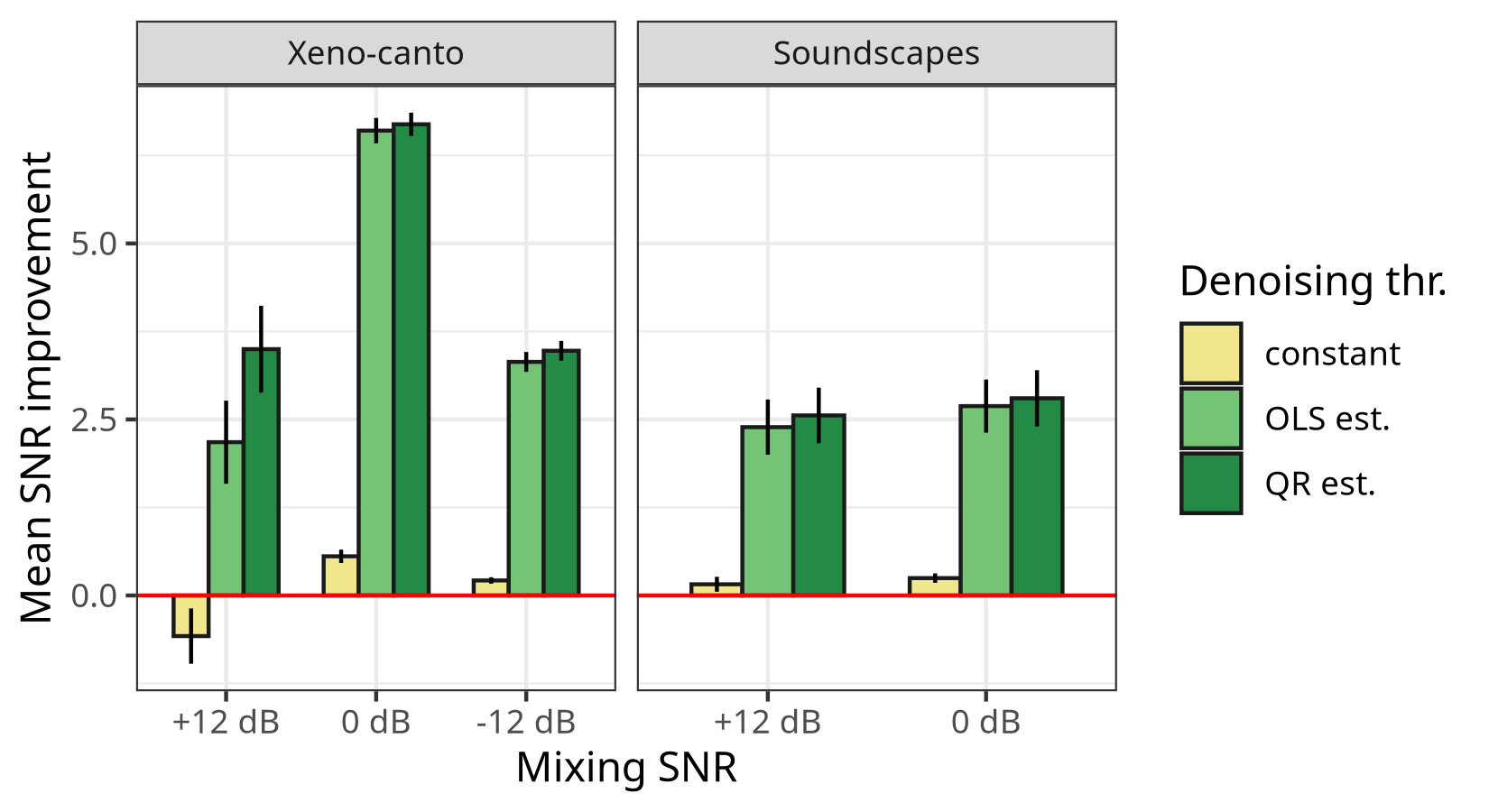}
  \caption{Average SNR improvement obtained with different wavelet denoising thresholds: constant, or based on the time-varying OLS or QR spectrum models, as presented here. Either high-SNR recordings of individual birds (xeno-canto) or dense soundscapes from passive monitoring data were mixed with wind noise at different SNRs, and denoised by wavelet shrinkage. Error bars show $\pm$ 1 SE.}
  \label{fig:dnbarplot}
\end{figure}

The proposed wind noise level estimators are useful for denoising as well.
Wavelet shrinkage with wind-adaptive thresholds, either estimated by OLS or by QR, considerably improved the SNR (Figure \ref{fig:dnbarplot}).
In contrast, the same denoising method with a constant threshold led to very little improvement: while it decreased the overall background noise, most of the noise energy in these examples came from wind gusts, which this method could not remove.
Note that in some cases SNR even apparently decreased: because some white noise was present in the ``clean'' recordings as well, removing that decreased the measured match between the reference and denoised files, and was thus counted as a loss of signal by this metric. (This effect also contributed to the lower denoising performance seen when using the soundscape references, which had more residual noise than the xeno-canto clips.)

Similar results are seen when using the SI-SDR metric (Supplementary Material, Figure \ref{fig:dnbarplotsupp}). As this metric is invariant to the initial mixing SNR, it produces more uniform measures over the tested files, removing the spurious improvement peak seen at 0 dB with the xeno-canto examples.
The difference between OLS and QR estimation methods was very small in the metrics used here, although in favour of QR in every case.

\begin{figure}[!h]
  \centering
  \includegraphics[width=0.8\textwidth]{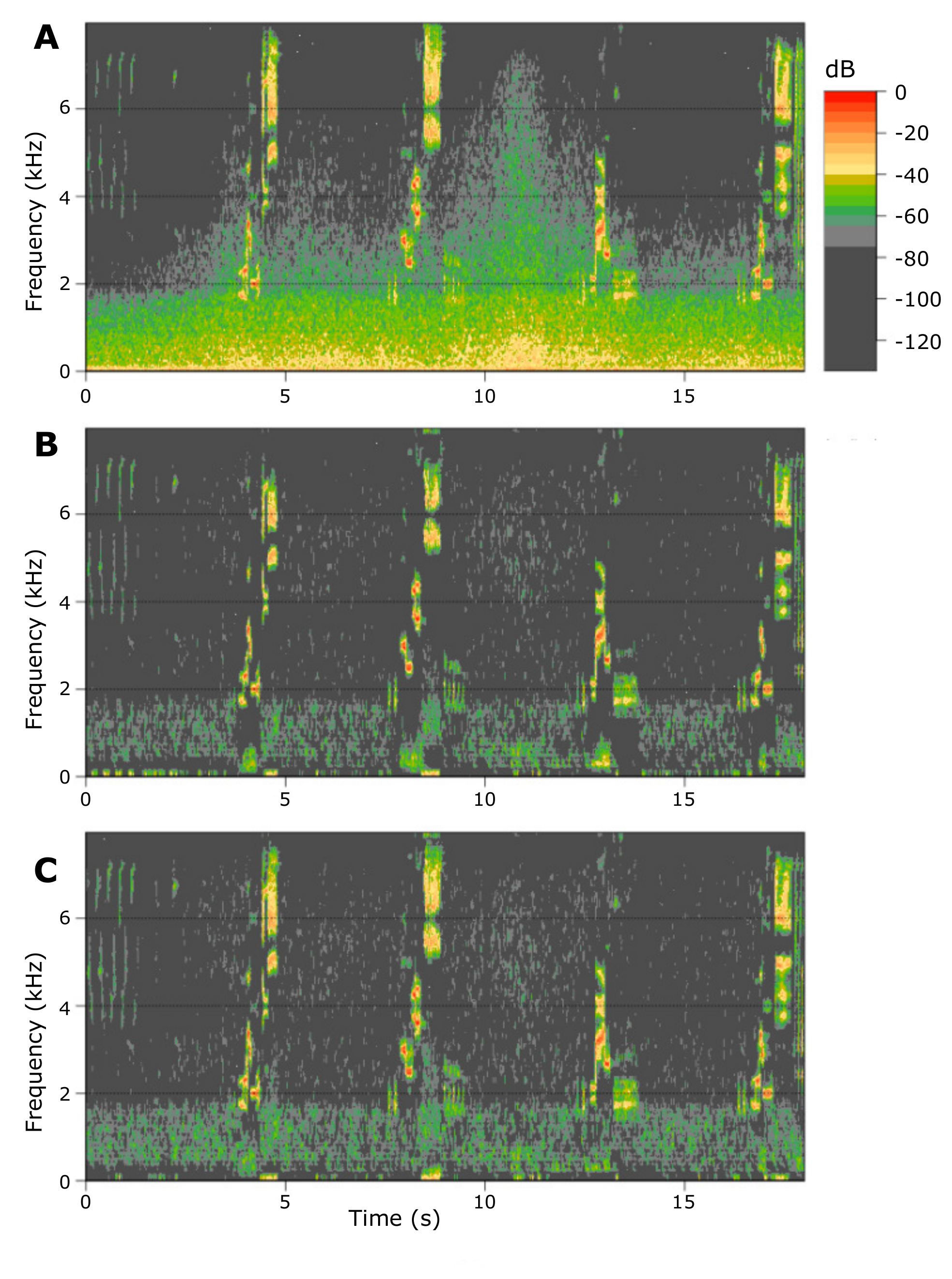}
  \caption{Recording of a wood thrush \textit{Hylocichla mustelina}, xeno-canto ID XC561864, mixed with wind noise at +12 dB, and denoised by wavelet shrinkage with different thresholds: (A) constant threshold, (B) adaptive threshold estimated by OLS, (C) adaptive threshold estimated by QR.}
  \label{fig:denoisedcalls}
\end{figure}

To gain some insight into the working of each method, we show spectrograms of the denoised outputs from one example clip in Figure \ref{fig:denoisedcalls}.
Denoising with a constant threshold successfully removed the more stable parts of the background noise, as indicated by uniform grey areas in Figure \ref{fig:denoisedcalls}A; however, it had very little effect on the wind gust seen around 10 s from the start of the clip.
Both of the time-varying estimators successfully modelled this gust, which led to its removal.
The main differences between the estimation by OLS and QR is seen during the time periods when loud calls are present: as predicted, these calls affect the OLS fit more, and cause over-adjustment (grey gaps) or under-adjustment (green residual noise) in the 0-2 kHz frequency range (Fig. \ref{fig:denoisedcalls}B). The QR estimate was robust to these effects (Fig. \ref{fig:denoisedcalls}C).

\section{Discussion}

\subsection{Summary and alternative design considerations}

In this study, we proposed a new noise estimator based on fitting a polynomial model to wavelet packet node energies.
This estimator was combined with log spectral subtraction to stabilize the noise level. In our case study this adjustment greatly reduced the number of false positive detections and led to more efficient acoustic surveys.
Additionally, we showed that the estimator can be incorporated in a wavelet denoising method to restore sound polluted by broadband noise.

Although our initial motivation was wind noise, which in theory is associated with a specific (log-)linear spectrum shape, the pilot experiment indicated that a more flexible model was needed.
The resulting polynomial estimator now also captures more general broadband noises besides wind.
This is useful for our surveys, but in other cases the target signal may be broadband, such as insect stridulations \citep{weta1992}.
Our method is thus limited to signals with characteristic frequency bands.
However, choosing other filterbanks instead of the wavelet packet, such as the Mel, gammatone or Greenwood \citep{Zeppelzauer2014}, may concentrate different sounds better, thus allowing analysis of a wide variety of tasks.

We proposed to fit the spectrum using quantile regression to account for asymmetric contamination when other signals are present.
Nonetheless, the standard OLS fitting appeared surprisingly robust, although choosing QR is safer when the soundscape is particularly rich, or precise noise estimation around calls is important.
This may be a useful precursor step for automatic analysis of dawn choruses, in which the high density of calls presents a challenge for current detection software \citep{Brooker2020}.

\subsection{Differences from other noise estimators}

The surveys analysed here highlight some of the issues with applying other noise estimation methods in bioacoustics.
Since our spectrum model uses short time windows (on the order of 0.1 s) and is not smoothed over time, it can adapt to fast transients, while methods such as PCEN \citep{pcen2019} or MMSE-STSA \citep{Brown2018} critically rely on the noise changing more slowly than the signal, so would be unusable with the c. 30-second-long kiwi call.
Methods designed to remove low-frequency noises, such as presented in \citet{NelkeCentr}, cannot be applied to the 150 Hz bittern sounds, and in general require knowledge of the other signals expected in the environment.
In contrast, the proposed noise estimator needs very little tuning to be applied to different species: the main parameter is the frequency range of the target, which is retrieved from the detection stage. Furthermore, it can often be used even without specifying the signal bands at all, as shown in the denoising examples.

Thus, the proposed framework is designed for low training data situations that are common in wildlife research, where recording collection and expert annotation is expensive.
In our survey analysis workflow, only the wavelet energy detector needs training, which has a simple structure and so can be trained with less than an hour of data \citep{wavelets2020, soundchps21}.
Neural networks could be used to create noise-robust detectors that outperform our results if given sufficient data, but this likely means at least thousands of clips, as in e.g., \citet{Vickers2021}.
Options to reduce this requirement, such as by transfer learning or using weak labels \citep{DCASE2018}, are actively researched.
Even with that, our method will still remain useful, as it provides a fast and robust initial screening step at very little cost in terms of missed calls, and its output can be verified by a more sophisticated procedure if available.

Additionally, we have used a wavelet transform throughout all stages of the sound analysis. This transform can be easily inverted, as we have done here to recreate the denoised audio files, whereas most other methods produce spectrograms, inverting which would be more complicated \citep{spinv2007}.


\subsection{Evaluating the improvements in practice}
The metrics chosen to evaluate the proposed methods may not represent all practical needs.
For the detection stage, we conducted a grid-based survey in the SCR framework and measured its efficiency as proposed in \citet{acousticsurveys}.
Alternative measures, such as the F-score, are common in the acoustic detection community (see e.g., \citet{wavelets2020}).
In contrast to these, our SCR metric directly measures the precision of the estimate of interest, and thus the power to conduct ecologically relevant comparisons.
To the best of our knowledge, this study is the first to explicitly demonstrate that survey efficiency is gained by using noise-robust sound analysis methods.
The metric is also quite general, as various bioacoustic survey designs can be expressed as special cases of SCR \citep{borchersunifying2015}.
Ultimately, in the present case, the robust detector showed much lower false alarm rate with almost no loss in true detections, so it should be identified as an improvement by most metrics.

It is yet more complicated to evaluate the benefits of denoising in bioacoustics.
If the cleaned sound is used for human listening, presence of perceptual artefacts such as musical noise may be more important than SNR \citep{VaseghiBook}.
Metrics such as PESQ have been designed to capture the subjective quality of sound \citep{pesq2001}, but they rely on speech-specific properties and do not directly apply to other species.
In the context of ecological monitoring, the primary application of denoising currently is to improve the classification of calls by neural networks, as in e.g., \citet{Vickers2021}.
Deep learning is also suggested for more holistic ecoacoustic assessments, outside traditional surveys, and removing noise is also of interest there \citep{Fairbrass2018}.
Because of the black-box nature of these methods and variety in the network and training setups, it is not clear whether SNR, PESQ or other metrics would actually be predictive of their performance.
Standardizing the protocols of training and applying neural networks in bioacoustics would allow one to investigate this relationship, and to further develop denoising methods that are beneficial in ecology practice.

\section*{Acknowledgements}

This research is supported by the New Zealand Marsden Fund, which is administered by the Royal Society of New Zealand Te Ap\=arangi under grants 17-MAU-154 and 17-UOA-295. We also thank Danielle Shanahan for the opportunity to work in Zealandia, and Alberto De Rosa for the Ponui field data.

\bibliography{wind-references.bib}

\clearpage
\setcounter{figure}{0}
\setcounter{subsection}{0}
\renewcommand{\thesubsection}{\Alph{subsection}}
\renewcommand{\thefigure}{S\arabic{figure}}

\section*{Supplementary Material}

\subsection{Estimation of quantiles with contamination}
\label{sec:contquantilesect}

Let $N$ be a random variable following a contaminated mixture distribution, i.e., its CDF is $F_N = (1-\epsilon)F_X + \epsilon F_Z$, 
with a contaminating variable $Z$ that is concentrated at larger values than $X$. Specifically, denoting the median of $X$ as $\mu_X$, define this requirement as:
\begin{equation}
   F_Z(x) \ll F_X(x) \text{ for all } x \le \mu_X.
  \label{eq:contamconcentr}
\end{equation}
Then its $\tau$-quantile for $\tau/(1-\epsilon)<0.5$ is:
\begin{align*}
  F_N^{-1}(\tau) &= \inf(x: F_N(x) \ge \tau) \\
    &= \inf(x: (1-\epsilon) F_X(x) + \epsilon F_Z(x) \ge \tau) \\
    & \approx \inf(x: (1-\epsilon) F_X(x) + 0 \ge \tau && \text { using \eqref{eq:contamconcentr} } \\
    &= \inf(x: F_X(x) \ge \tau/(1-\epsilon)) \\
    & = F_X^{-1}(\tau/(1-\epsilon))
\end{align*}
The $N$, $X$ and $Z$ variables correspond to $|N_{j,k,t}|^2$, $c(t)f_{j,k}^{-\alpha}\chi^2_1$ and $Z_{j,k,t}$ from \eqref{eq:contamdistr}.
Thus, the quantile regression estimate $\hat{N}^{Q,\tau}$, which converges to $F_N^{-1}(\tau)$ under standard regression conditions, also converges to the $\tau/(1-\epsilon)$ quantile of the ``clean'' distribution $c(t)f_{j,k}^{-\alpha} \chi^2_1$.

Naturally, neither $\epsilon$ nor the exact range of $\tau$ over which the requirement \eqref{eq:contamconcentr} holds are known in advance. However, choosing a small quantile such as $\tau=0.2$ should work for most situations in practice: even under 50 \% contamination, this corresponds to $\tau/(1-\epsilon) = 0.4 < \mu_X$, and we can reasonably expect that most signals will significantly exceed the median of the background.

\subsection{Tail probabilities of spectral subtraction estimates}
\label{sec:spsubtailsect}
Let $X^2_t$ be a random process with $X^2_t \sim c(t) \chi^2_1$, where the noise strength $c(t)$ varies with time.
Assume the expected value of the process at each time $t$, i.e., $\mathbb{E}X^2_t = c(t)$ is known exactly.
We wish to find the tail probabilities of $\tilde{X}^2_t$, obtained by standard (power) spectral subtraction \eqref{eq:spsub}:
\[ \tilde{X}^2_t = \max(0, X^2_t - c(t)). \]
This is by definition a left-censored variable, with pdf
\[ f_{\tilde{X}^2_t}(x) = \begin{cases}
    f_{X^2_t - c(t)}(x) = f_{X^2_t}(x+c(t)) & \text{ if } x>0 \\
    F_{X^2_t - c(t)}(0) & \text{ if } x=0.
  \end{cases}
\]

The tail probability for any $\lambda > 0$ is
\begin{align*}
  P(\tilde{X}^2_t > \lambda) &= \int_{\lambda}^\infty f_{X^2_t}(u+c(t)) du \\
    &= \int_{\lambda+c(t)}^\infty f_{X^2_t}(x) dx \\
    &= 1- F_{X^2_t}(\lambda+c(t)) \\
    &= 1- F_{\chi^2_1}\left( \frac{\lambda+c(t)}{c(t)} \right) .
\end{align*}
Note that in the denoising context, this ``denoised'' distribution still depends on the noise strength $c(t)$.

Under log spectral subtraction, we immediately have that $X^2_t/c(t) \sim c(t)/c(t) \chi^2_1 = \chi^2_1$, and thus the tail probabilities for any $\lambda>1$ are $1-F_{\chi^2_1}(\lambda)$. The rectification $\max(1, \cdot)$ can be omitted if the distribution properties for $\lambda<1$ are also relevant.

\clearpage
\subsection{Supplementary Figures}

\begin{figure}[h]
  \centering
  \includegraphics[width=\textwidth]{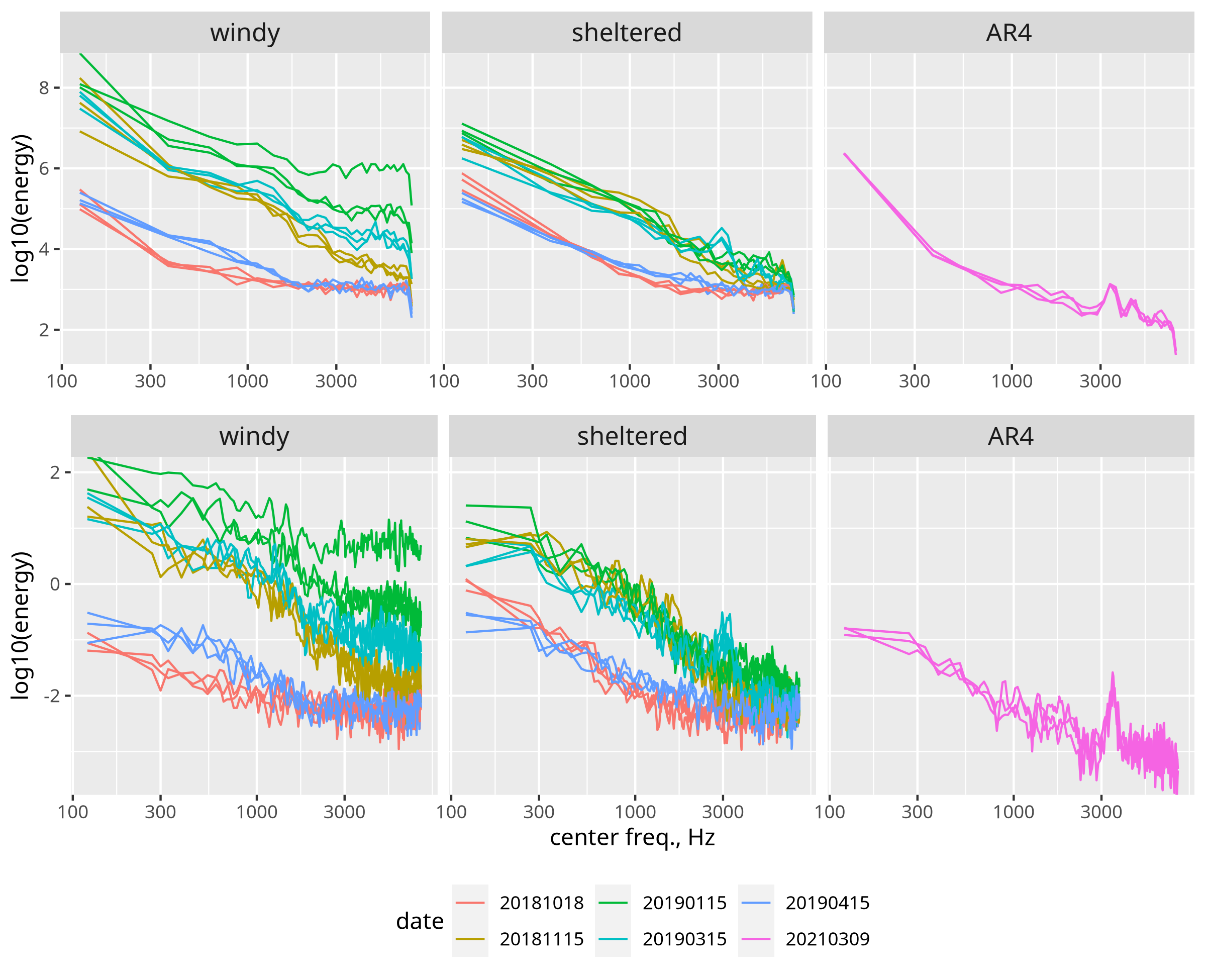}
  \caption{(Top row) Short-term spectra obtained using a Symlet(8) wavelet packet on 0.1 s clips from passive acoustic recordings. The log energy of each node is plotted against its centre frequency. Panels show clips from a recorder in a windy location, a sheltered location, and a different model of recorder (AR4). (Bottom row) Short-term spectra of the same clips estimated by periodogram, with Daniell smoothing over 7 bins. Note the considerably higher variance of this estimator, compared to wavelet packet spectra.}
  \label{fig:pilotressupp}
\end{figure}

\begin{figure}[h]
  \centering
  \includegraphics[width=0.8\textwidth]{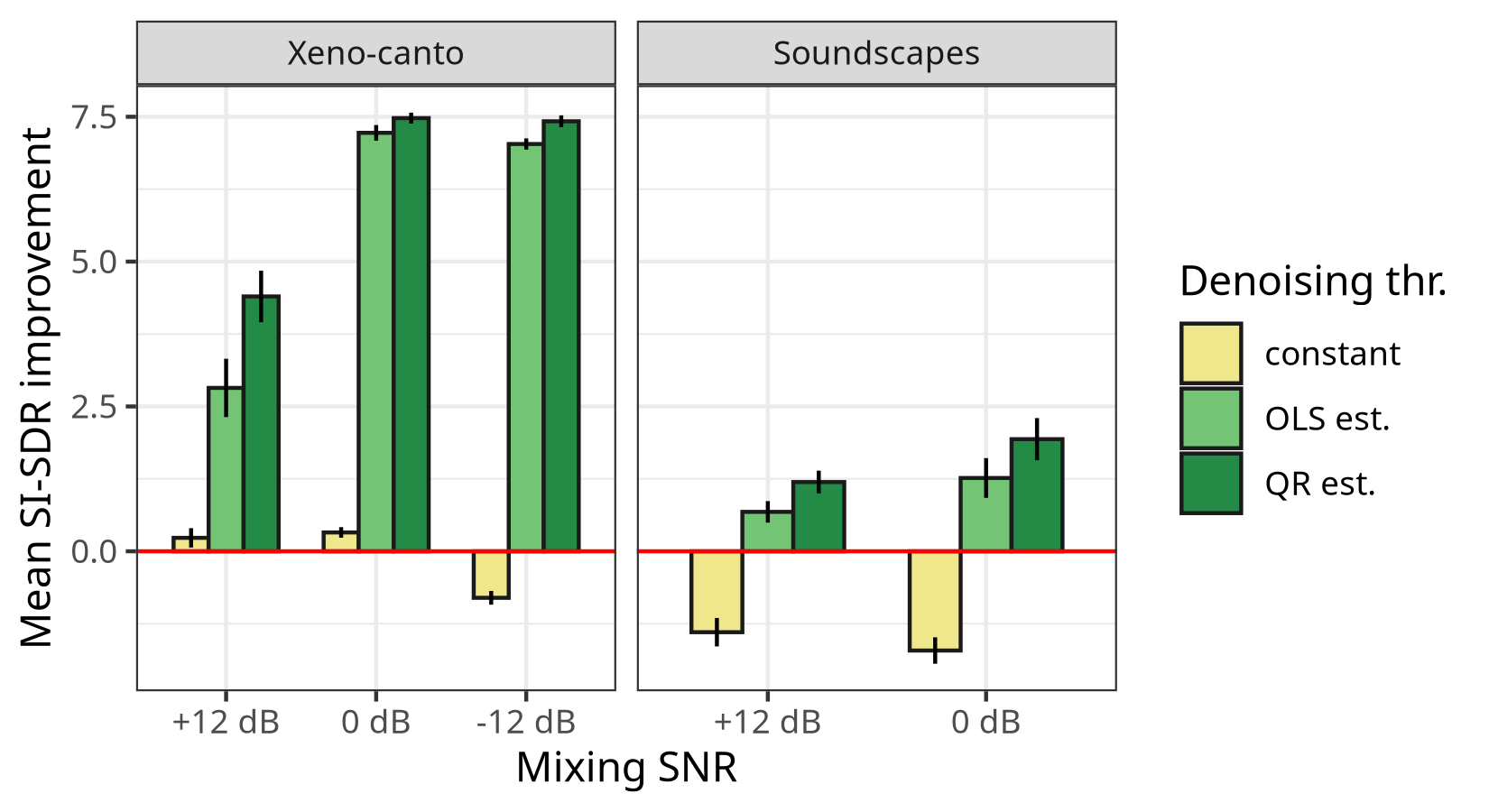}
  \caption{Average improvement in scale-invariant SDR obtained with different wavelet denoising thresholds: constant, or based on the time-varying OLS or QR spectrum models, as presented here. Either high-SNR recordings of individual birds (xeno-canto) or dense soundscapes from passive monitoring data were mixed with wind noise at different SNRs, and denoised by wavelet shrinkage. Error bars show $\pm$ 1 SE.}
  \label{fig:dnbarplotsupp}
\end{figure}

\end{document}